\documentclass[12pt]{iopart}

\usepackage{graphicx}
\begin{document}

\title[]{ A comparative DFT study of electronic properties of 2H-, 4H- and 6H-SiC($0001$) and SiC($000 \overline{1}$) clean surfaces: Significance of the surface Stark effect}

\author{Jakub So\l{}tys$^1$, Jacek Piechota$^1$, Micha\l{} \L{}opuszy\'nski$^1$ and Stanis\l{}aw Krukowski$^2$}
\address{$^1$Interdisciplinary Centre for Materials Modelling, University of Warsaw,\\ Pawi\'nskiego 5a, 02-106 Warsaw, Poland \\
$^2$Institute of High Pressure Physics, Polish Academy of Sciences, Soko\l{}owska 29/37, 01-142 Warsaw, Poland
 }

\begin{abstract}
Electric field, uniform within the slab, emerging due to Fermi level pinning at its both sides is analyzed using DFT simulations 
of the SiC surface slabs of different thickness. It is shown that for thicker slab the field is nonuniform 
and this fact is related to the surface state charge. Using the electron density and potential profiles 
it is proved that for high precision simulations it is necessary to take into account enough number of the Si-C layers.
We show that using 12 diatomic layers leads to satisfactory results.
It is also demonstrated that the change of the opposite side slab termination, both by different type 
of atoms or by their location, can be used to adjust electric field within the slab, creating 
a tool for simulation of surface properties, depending on the doping in the bulk of semiconductor. 
Using these simulations it was found that, depending on the electric field, the energy of the surface 
states changes in a different way than energy of the bulk states. 
This criterion can be used to distinguish Shockley and Tamm surface states.
The electronic properties, i.e. energy and type of surface states of the three clean 
surfaces: 2H-, 4H-,  6H-SiC($0001$), and SiC($000 \overline{1}$) 
are analyzed and compared using field dependent DFT simulations. 

\end{abstract}
\maketitle

\section{\label{sec:level1}Introduction }
Silicon carbide can exist in very large number of polytypes, of which hundreds were already identified \cite{Kackell}. 
They are created due to different stacking sequences, still preserving tetrahedral configuration of silicon and carbon atoms. 
These polytypes belong to wide bandgap semiconductors, having its gaps, ranging from 2.403 eV (3C), 3.101 eV (6H), 
3.285 eV (4H) to 3.300 eV (2H) \cite{Masri}. Also, SiC stacking faults, being a planar large area defects, could be obtained in a controlled fashion. 
These structures, chemically identical with the material of the locally adjustable bandgap, are new, potentially important, quantum structures \cite{Bechstedt}. 
In addition, a number of quantum structure based devices, using SiC as building material was developed \cite{Bechstedt}. 
These structures are based on the material having its properties, extremely attractive in many applications in electronics \cite{ru}, 
that is a wide bandgap semiconductor, having superior electronic properties, displaying high values of several important parameters 
like breakdown electric field ($2.2 \cdot 10^6 \; \rm{V} \cdot \rm{cm}^{-1}$), forward current density (up to $1 \; \rm{kA} 
\cdot \rm{cm}^{-2}$),  saturated electron drift velocity 
($2 \cdot 10^7 \; \rm{cm} \cdot \rm{s}^{-1}$ for 4H-SiC and $2.5 \cdot 10^7 \; \rm{cm} \cdot \rm{s}^{-1}$ for 3C-SiC) and room 
temperature mobility ($370 \; \rm{cm}^2 \cdot  V^{-1} \cdot s^{-1}$ and $720 \; \rm{cm}^2 \cdot \rm{V}^{-1} \cdot \rm{s}^{-1}$ 
for 6H-SiC and 4H-SiC respectively), and finally, high blocking voltage observed in typical MOSFET devices \cite{Bechstedt}. 
It may be used in high temperature environment because it is thermally resilient and has high thermal conductivity \cite{Tang}. 
Being mechanically extremely hard and chemically inert \cite{Goldberg}, SiC can be applied in a number of important applications, 
mostly in high power, high temperature and high frequency electronics \cite{Capano}. 
Therefore SiC and the structure based on this compound attracted tremendous interest resulting in huge number of publications. 

The most successful method of the growth of large size high quality crystals is based on physical vapor transport (PVT), 
known as a modified Lely method first developed by Tairov and Tsvetkov \cite{Tairov1,Tairov2}. The method, though technically 
difficult due to very high temperatures required, was developed to the stage in which the large size SiC crystals are grown 
and are distributed commercially. Yet, despite considerable success, the method is still plagued by difficulties most 
notably by creation of micro- and nano-pipes \cite{Verma}, that deteriorate or even block functioning of SiC-based devices. 
The density of these defects was substantially reduced which allowed to use of Lely grown SiC substrates in 
technology of electronic devices. The pipe problem still remains essential, thus basic research and development of 
the technology are both needed to alleviate it completely. 

It is well known that the modified Lely method relies on the chemical decomposition of the source SiC material into Si, 
$\rm{Si}_2\rm{C}$ and $\rm{SiC}_2$ volatile species. The transport process is, therefore, prone to the chemical instabilities, 
which affect the 
chemical state of SiC surface, leading to structural and chemical transformation which may contribute to the generation 
and preservation of the pipes during the growth. It is therefore understandable that the atomistic structures of various 
SiC surfaces, belonging to different polytypes, were intensively studied. Here we limit our discussion to the stoichiometric 
polar surfaces of hexagonal polytypes, i.e. 2H-, 4H- and 6H-SiC($0001$) and SiC($000 \overline{1}$) surfaces. 

In fact, the clean hexagonal SiC (0001) surfaces are difficult to achieve, as they are typically covered by Si or C atoms, 
or other adsorbed species like oxygen, hydrogen etc. Despite the fact that the DFT studies of these surfaces have considerable value 
as the reference point for the investigations of the adsorption phenomena and their influence on the electronic and structural 
properties of SiC($0001$) and SiC($000 \overline{1}$) surfaces. In an early study of these systems, Sabisch et al. \cite{Sabish} showed that both Si- and 
C-terminated surface undergo strong relaxation of the topmost layer by surface reconstruction. 
The top Si and C layers move 0.15 \AA \ and 0.25 \AA \ towards the SiC interior, a considerable fraction (about $23\%$ and $40\%$) of 
ideal 0.63 \AA \ distance. The second layer, i.e. C and Si layers move in opposite direction by 0.04 \AA \ and 0.11 \AA \, 
which is much smaller fraction of the ideal distance 1.88 \AA. Third layer relaxation is small, moving the atoms by 0.02 \AA \ and 0.03 \AA, respectively. 
Again the direction is reversed, the third layer atoms are moved towards SiC interior. 

The above described relaxation type indicates that the ionic contribution in this effect is essential and the 
subsurface fields are important factors. Despite that these authors calculated surface band structure of both surfaces 
without reference to this field \cite{Sabish}. Using DFT LDA, they obtained indirect bandgap of 6H-SiC bulk equal to 1.97 eV, 
which reflects well known under-estimate of bandgap energy in this approach. The band structures of both surfaces were 
compared with the bulk data. They identified the flat band in the gap, originating from dangling bonds of surface Si and C atoms. 
The Si related surface state is about 1.5 eV above the C-related states which was attributed to difference in ionic character of both surfaces. 
A charge transfer towards C-surface was predicted and proved by the charge density distribution plots. 
It was also shown that both states are half-filled i.e. the surface is metallic. It is interesting to note that 
the charge density distribution of the dangling bonds is totally different: C-state is quasi-symmetric with some 
dominance of the internal part, while in Si-related state the charge is located exclusively in the region above the Si atom. 
The dispersion of both states is generally small which is due to small overlap of the neighboring atom wavefunctions.
As expected, larger dispersion is observed for Si-related states \cite{Sabish}. 

The adsorbate-covered SiC($0001$) and SiC($000 \overline{1}$) surfaces display large variety of different reconstructions such as 
($\sqrt{3} \times \sqrt{3}$), ($3 \times 3$), ($6 \sqrt{3} \times 6 \sqrt{3}$) and ($9 \times 9$) \cite{Catellani}. The plethora of different symmetries was studied 
and several different models were proposed \cite{Starke,Kulakov,Li}. The subject remains under intense dispute. A new impetus to the 
studies of nonstoichiometric SiC surface was given by the discovery of graphene on both Si- and C-terminated SiC surfaces \cite{Hass}. 
Since the subject is vast and not in the mainstream of the present work it will not be discussed here. 

It is interesting to note that the surface structure, defined by the positions of the adsorbed atoms is very 
sensitive to external factors. This effect was investigated by Righi et al. \cite{Righi} They have found that, at the surface, 
the two-dimensional effect favor the formation of 3C-SiC. In contrast to adsorbate covered surfaces, the clean polar 
surfaces of other hexagonal polytypes were expected to have essentially similar properties, as the differences in 
atom arrangement are observed deep in the bulk.  Still this effect could be investigated along with the role of the subsurface electric 
field. 
\begin{figure}
\begin{center}
\includegraphics[width=0.7 \textwidth]{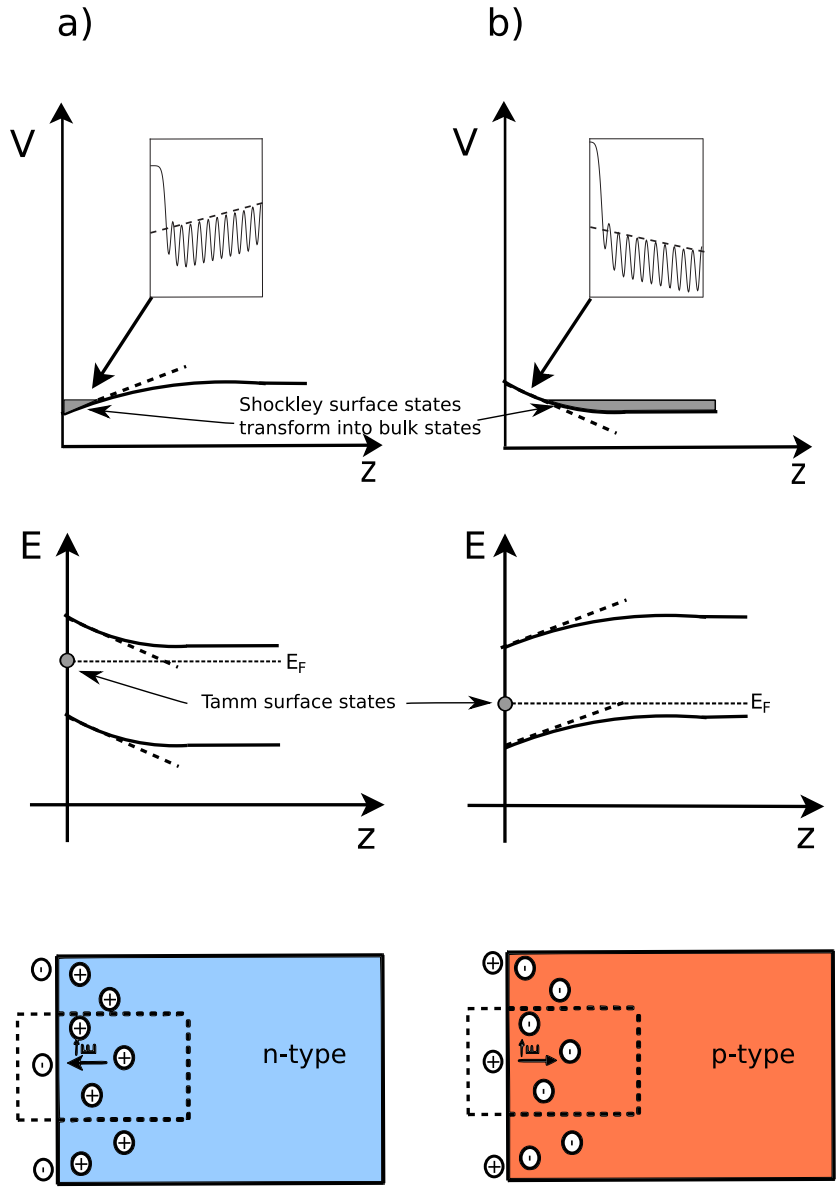}
\end{center}
\caption{\label{fig:acpdon} (Color online) Electrostatic potential at the surface edge (top) , 
band structure (middle) presented for n-type and p-type semiconductor (bottom).\\
For n-type semiconductor Shockley surface states are located on the edge of surface, 
for p-type they are shifted to the bulk, i.e. these surface states disappear. 
Unlike Shockley, Tamm surface states appear in the band gap and can move across it. 
}
\end{figure}
 
The present paper is devoted to study the methodology of thick slab simulations, 
the role of electric field (Fig. \ref{fig:acpdon}) and the influence of stacking sequence on electronic and structural properties  
of polar SiC($0001$) and SiC($000\overline{1}$) surfaces. 
The paper is organized as follows. In Sec. \ref{Sec:Method} we give a brief overview of employed computational methods.
Sec. \ref{Sec:SurfaceModels} describes the construction of surface models. We analyze there the influence of
finite number of layers on reliability of calculations. We also study the effect of surface termination by hydrogen atoms
and show that they can be used to adjust electric field inside a slab.
Fig. \ref{fig:acpdon} explains the subsurface field and its connection with surface acceptor 
and donor states together with its modeling in the slab simulations. 
As described in detail in Sec. \ref{Sec:SurfaceProperties}, there is significant difference between Shockley and Tamm surface states. 
As showed in Fig. \ref{fig:acpdon},  Shockley states can be transformed into bulk states under sertain electric field conditions. 
On the other hand, Tamm states, irrespective of electirc field, are placed in the band gap.  
In Sec. \ref{Sec:SurfaceProperties} we present obtained structural and electronic properties of examined surfaces.
Sec. \ref{Sec:Summary} summarizes our work.

\section{Method of calculation \label{Sec:Method}}

We have carried out self-consistent total-energy calculations based on density functional theory.
The code VASP developed at the Institut f\"ur Materialphysik of Universit\"at Wien was employed
\cite{Kresse1993,Kresse1996,Kresse1996a}. The projector augmented wave (PAW) approach \cite{Blochl} 
was used in its variant available in the VASP package \cite{Kresse1999}.
For the exchange-correlation functional generalized gradient approximation (GGA) according to Perdew, Burke and 
Ernzerhof (PBE) \cite{Perdew1996} was selected.
A plane wave cutoff energy was set to 500 eV.
In bulk calculations 11x11x11 Monkhorst-Pack k-point mesh was used \cite{Monkhorst}. 

\begin{figure}
\includegraphics[width= \textwidth]{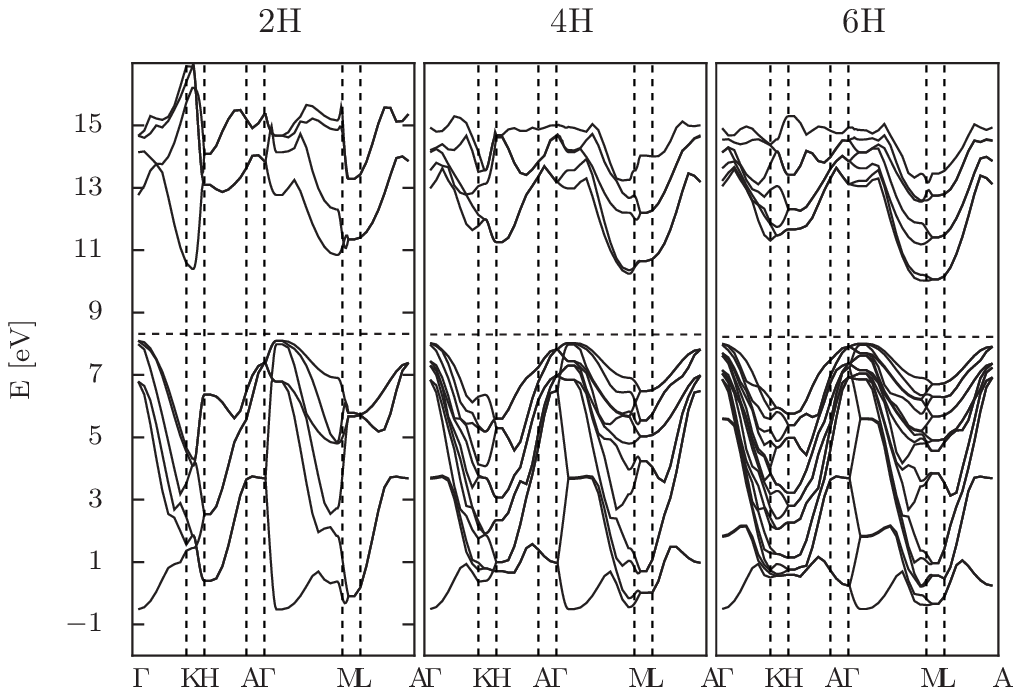}
\caption{\label{fig:bulk} Band structure calculated using selected parametrization for the three hexagonal polytypes: 2H-, 4H- and 6H-SiC($0001$).
Horizontal dotted line shows the Fermi level. }
\end{figure}

Lattice parameters obtained for bulk SiC polytypes (2H, 4H, 6H) summarized in Table  \ref{tab:table1}
are in good agreement with the experiment \cite{ru,Goldberg,Schulz}.
\begin{table}
\caption{\label{tab:table1} 
Lattice constants and bandgaps of bulk 2H-, 4H- and 6H-SiC (DFT results and experimental data \cite{ru,Goldberg,Schulz}).
Relative deviations from experiment are given in parentheses.  }
\begin{center}
\begin{tabular}{cccc}
Crystal & 2H & 4H & 6H \\
\hline
a [\AA] - DFT & 3.092 (0.4\%) & 3.094 (0.7\%) & 3.095 (0.7\%) \\
c [\AA] - DFT & 5.074 (0.4\%) & 10.129 (0.8\%) & 15.187 (0.5\%) \\
\hline
a [\AA] - exp  & 3.079 & 3.073 & 3.073 \\
c [\AA] - exp  &  5.053 & 10.053 & 15.118 \\
\hline
$E_g$ [eV] - DFT & 2.29  & 2.21  & 2.03  \\
$E_g$ [eV] - exp & 3.33  & 3.285  & 3.101  \\
\end{tabular}
\end{center}
\end{table}
The band structures of the hexagonal polytypes are presented in Fig.~\ref{fig:bulk}. Generally these band structures are 
in good agreement with the previously obtained DFT results \cite{Sabish}.
As it follows from these data, the bandgaps are much lower than the experimental findings. 
This is a well known systematic error of DFT LDA/GGA approximations.

The surface slabs representing the 2H-, 4H- and 6H-SiC($0001$) surfaces are built using 
different number of Si-C double layers. We used standard notation i.e. SiC (0001) 
surface is the one terminated by Si atoms. The slabs having 6-18 atomic layers 
were used, i.e. the slabs consisting of up to 36 atoms.
Due to the fact that SiC(0001) and  SiC(000-1) surfaces can be prepared without surface reconstruction \cite{Emtsev},
(1x1) supercells were employed. 
The bottommost C atoms had their broken bonds either terminated by hydrogen atoms, 
or left uncovered. A particular attention was paid to k-point sampling and convergence. 
In surface calculations Monkhorst-Pack scheme was applied, using 9x9x1 k-point mesh
for bare and covered surfaces. 
Selected parameters were sufficient to obtain relative energies converged
to few meV.
To prevent artificial electrostatic field in vacuum region between repeated unit 
cells a dipole correction was used \cite{Neugebauer}. 
Three topmost Si-C layers were relaxed using a conjugate-gradient (CG) algorithm in Sec. 4.1 (Structural properties). 
Due to the fact that significant relaxation is limited to the first two layers in other cases only two layers were relaxed.

\section{Preparation of surface models \label{Sec:SurfaceModels}}
Recently developed formalism allowing to simulate the influence of the electric fields and the 
doping of bulk semiconductor allows to study the interplay between the ionicity of the surface, the existence of electric 
field and the position of Fermi level \cite{Krukowski1,Krukowski2}. 

\begin{figure}
\includegraphics[width=\textwidth]{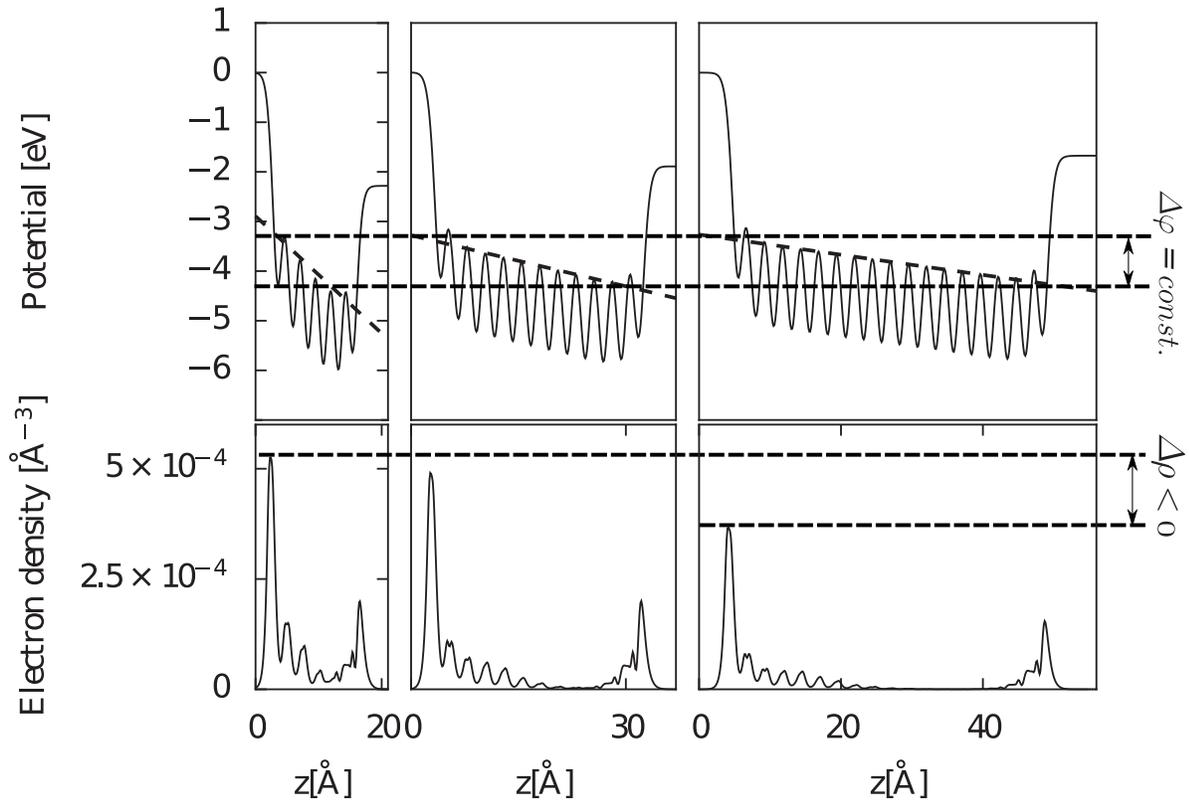}
\caption{\label{fig:field2h} Surface charge density and electric field in the 2H-SiC slab, 
obtained from DFT calculations for different thickness: 6 Double Layers (DL)  left, 
12 DL middle and 18 DL  right. Potential profile along channeling path (top), charge 
density associated with the Si and C surface states (bottom).  }
\end{figure}

Mayer and Marx \cite{Meyer} claimed that ZnO surface can be described using purely ionic model, 
where each Zn-O double layer exhibits a dipole moment perpendicular to the surface. In such model we obtain 
spontaneous polarization which is independent of the thickness of the slab.
Top part of Fig. \ref{fig:field2h} shows that potential difference at both sides 
is independent of slab thickness, which means that field inside the slab 
is smaller for larger thickness and there is divergence for large thickness. Thus Mayer and Marx prediction is incorrect. 
Moreover, electron density at the ends of the slab decreases ( $\Delta \rho < 0$) for thicker surface. 

The energy of quantum states of H termination atom can be controlled by their distance to the nearest Si/C atoms. 
In accordance to the stability condition of fermion systems, the Fermi energy should be positioned  halfway 
between the empty and occupied states. The energy of the surface states may be changed by electric field 
which is well known Stark effect. Meyer and Marx \cite{Meyer} recognized that fact, but they erroneously assumed  
that these are always conduction and valence band states. In fact, more frequently these are surface states. 
Also following Tasker \cite{Tasker}, they deduced that the field in slab is constant which leads to divergence for large slab thickness. 
This was repeated later by Noguera \cite{Noguera}. This prediction also turned out to be incorrect. In fact, the condition of 
the stability of fermion system leads to smaller field in thicker slab as it was shown in Fig. \ref{fig:field2h}. 
From these calculations we can learn about influence of subsurface electric field on the equilibrium and dynamic 
properties of surfaces. This in turn depends on the energy and concentration of principal point defects in the bulk. 
At present this is completely unexplored area.

Electric potential energy profiles are useful quantities to examine 
when analyzing the influence of the finite size
of the slab. As it follows from Fig. \ref{fig:field2h}, these 
profiles inside the slab consist of three parts: the linear 
part in the middle, and nonlinear parts at both ends. The linear part is in 
accordance with the scenario predicted first by Meyer and Marx \cite{Meyer}, where they 
claimed that the electric field is created due to charge transfer within 
the slab leading to the equality of the Valence Band Maximum (VBM) on 
the negative charge side and Conduction Band Minimum (CBM) at the Fermi 
energy level. Using ionic arguments in investigation of ZnO surfaces, 
they claimed that the charge separation leads to uniform field within the slab, 
as is observed in the central part of potential diagrams in Fig. \ref{fig:field2h}. 
This argument was based on earlier argument by Tasker \cite{Tasker} that charge separation 
creates uniform field within the slab which leads to divergence for thicker slabs. 
Later on this argument was reformulated by Noguera for the case of oxide crystals  \cite{Noguera}. 
As it was discussed in Ref. \cite{Krukowski2}, this prediction is incorrect, the field is not kept 
constant but the difference in electric potential between both sides of the slab is 
preserved. This is also demonstrated in the potential plots in Fig. \ref{fig:field2h}, where 
the difference of potentials is identical for all three slabs. Thus for thicker slab, 
the electric field is smaller. 
\begin{figure*}
\includegraphics[width= \textwidth]{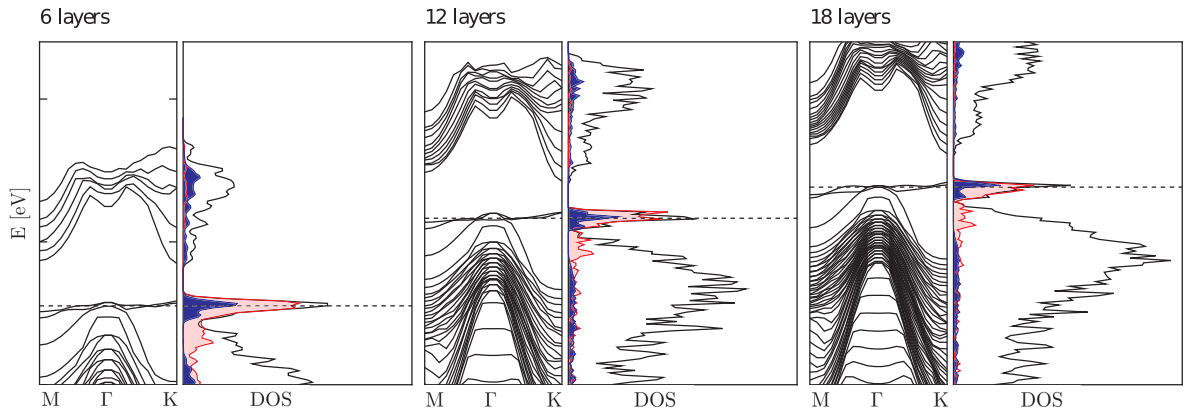}
\caption{\label{fig:banddos2h} (Color online) Band diagram (BD) and density of states (DOS) for three 2H-SiC slabs 
for different thickness: 6 DL  left, 12 DL middle and 18 DL right. 
The states associated with the surface Si(top) and C(bottom) atoms, multiplied by two, as well as total density of states 
are presented using different colours: blue, red, black, respectively. Dotted line shows the Fermi level.}
\end{figure*}

\begin{figure}
\includegraphics[width=\textwidth]{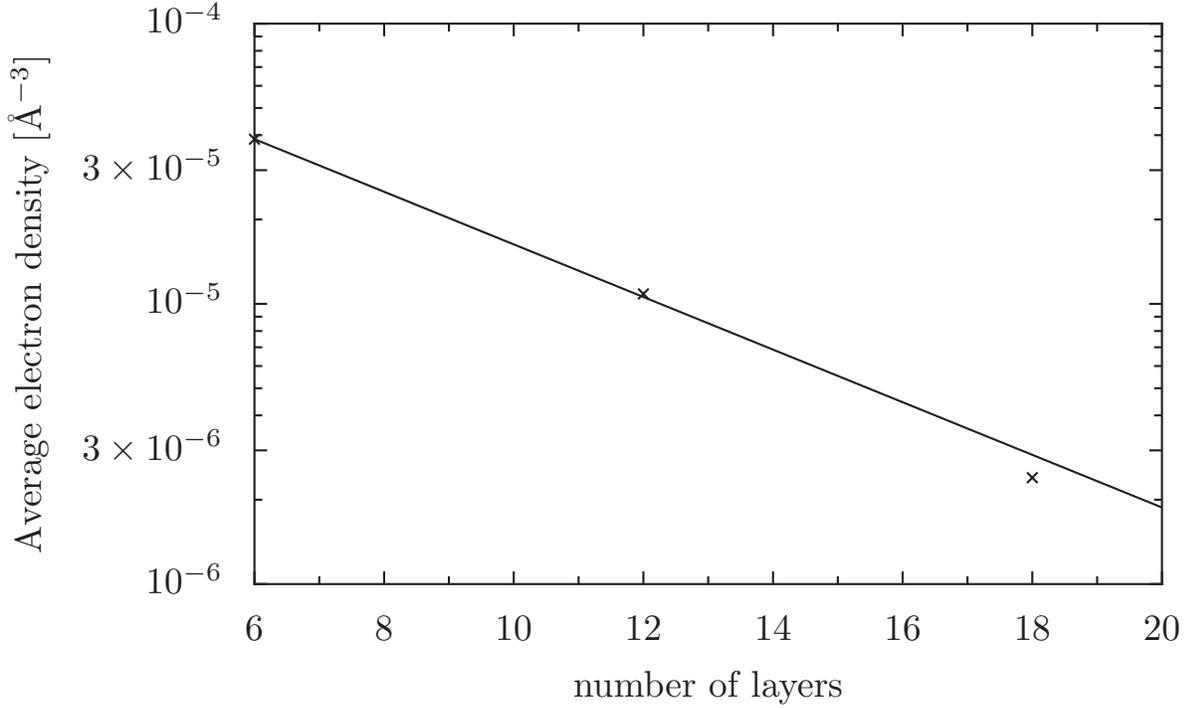}
\caption{\label{fig:surf2h} Charge distribution for different number of layers 
(6 DL, 12 DL and 18 DL ) and average electron density in the centre of the slab.
Here average charge is defined as average charge in the center of slab, i.e., 
interval with a width of 2.5 \AA \  from the center of the slab.}
\end{figure}

The identification of the physical factor creating the field within the slab 
by Meyer and Marx \cite{Meyer} is confirmed by the band diagram (BD) and density of states (DOS), 
plotted in Fig. \ref{fig:banddos2h}. 
These BD and DOS are calculated for the same slab models as electric field 
and charge diagrams presented in Fig \ref{fig:field2h}.
As it is shown for all these three cases, the Fermi level is kept 
at the energy of the quantum  states associated with both sides of the slab. 
Our BD and DOS prove that the Fermi level is 
attached to the states, independently on the slab thickness. 
Thus the diagram proves that the Fermi level is pinned 
at the bare Si-face and C-face.
The data presented in Fig. \ref{fig:field2h} show that the central part of 
the potential diagram is flat, i.e., this part of the slab is electrically neutral. 
On the contrary, the edge parts are not, which proves that there is an uncompensated charge at both 
sides of the slab, creating the uniform field between. This charge was plotted 
in Fig. \ref{fig:field2h} for systematic variation of the slab thickness and it was also shown  
that average charge density decreases exponentially (Fig. \ref{fig:surf2h}). Here average charge is 
defined as average charge in the center of slab, i.e., interval with a width of 2.5 \AA \  from the center of the slab.
In fact, the interaction 
between both sides of the slab have two principal contributions: electrostatic 
and quantum overlap. Electrostatic contribution is in fact useful for simulation 
of subsurface fields. The quantum overlap should be avoided as it could cause 
spurious effects. Hence both diagrams can be used for the determination how thick 
slab should be used for simulation of SiC surfaces. The linear part in potential 
profile and the negligible overlap (Fig. \ref{fig:surf2h}) are the necessary conditions
for accurate modelling. From these 
diagrams it follows that the slab having 12 DL is sufficient to simulate SiC polar surfaces. 

An electric field is an additional important component affecting the structure 
and the electronic properties of the semiconductor surfaces. The extent of 
this modification needs direct modelling by numerical procedures. 
According to our best knowledge, no such awareness or the methodology can be found in the literature. 
However, suitable approach was developed lately \cite{Krukowski1,Krukowski2}.
It was demonstrated that the change of 
the opposite slab termination can be used for simulation of the subsurface field 
in polar GaN(0001) surfaces. It was also shown that the change of the distance between 
the hydrogen termination atoms and the layer can affect the electric field, allowing 
to change its magnitude and also its direction. 
\begin{figure}
\begin{center}
\includegraphics[width=0.7 \textwidth]{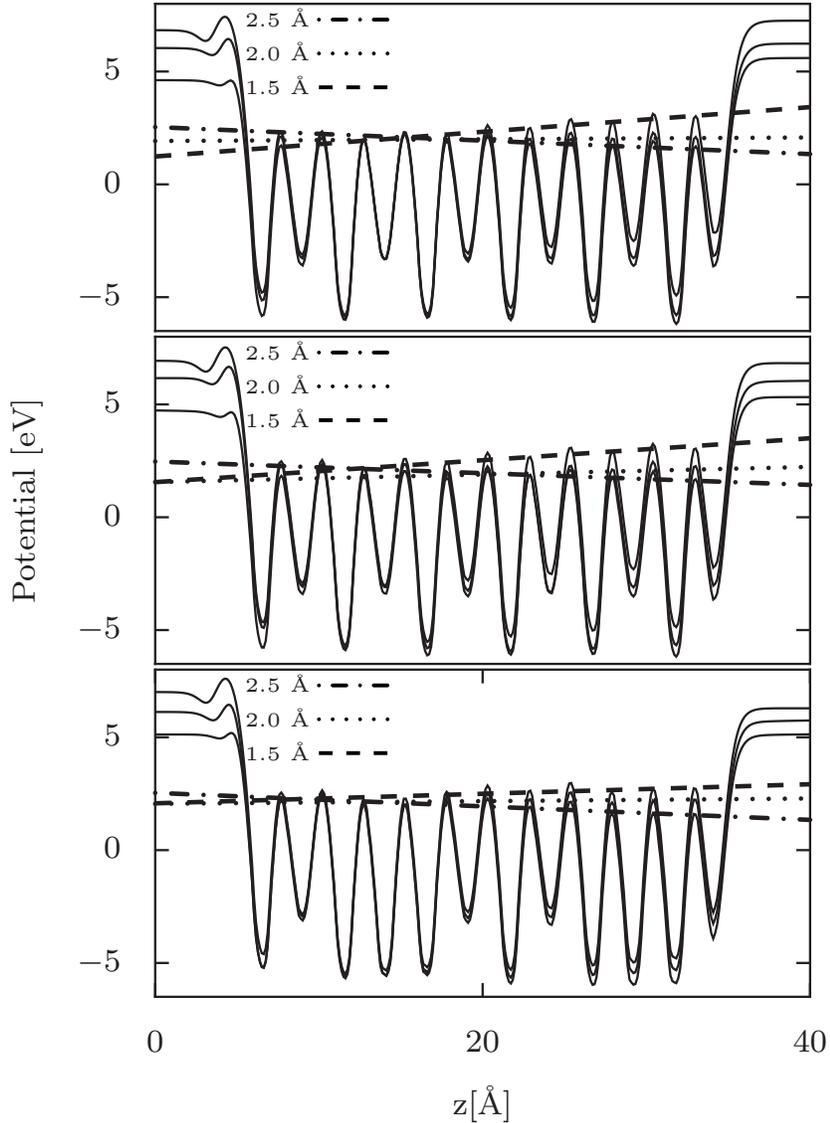}
\end{center}
\caption{\label{fig:fieldwithh2h} 
The electric field obtained from DFT calculations for 
different termination by H-atoms: 2H (top), 4H (middle) and 6H (bottom). 
Hydrogen atoms were placed at 1.5 \AA, 2.0 \AA, and  2.5 \AA \ from the bottom of SiC($0001$) surface. 
}
\end{figure}
The same approach was applied to SiC slabs. In Fig. \ref{fig:fieldwithh2h} the results of such 
approach are shown. It is demonstrated that the approach is universal: 
in all three polytypes the procedure allows us to obtain different 
magnitude and direction of electric field within the slab.
\begin{figure}
\includegraphics[width=\textwidth]{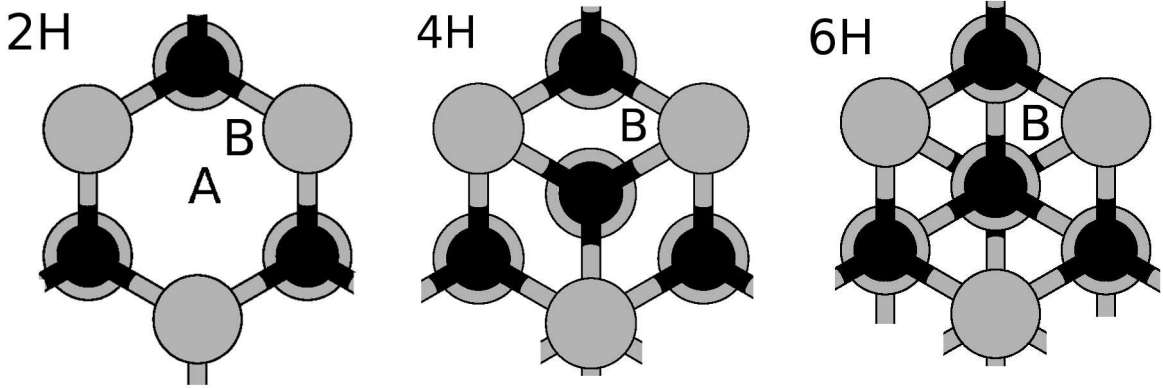}
\caption{\label{fig:fieldpoints} 
Top view of 2H-, 4H- and 6H-SiC($0001$) (Si- gray, C- black atoms). Points in which we get potential profiles 
are marked. 
}
\end{figure}
In Fig. \ref{fig:fieldpoints}  we show points in which we get potential cross sections. 
Potential profiles presented in Fig. \ref{fig:field2h} (2H-SiC) comes from point A in the middle of hexagon. 
4H- and 6H-SiC polytypes have atoms in point A. In this case point B is used. 
In order to achieve consistency with other polytypes, point B was used for each potential profile in Fig. \ref{fig:fieldwithh2h}. 

\section{Surface properties \label{Sec:SurfaceProperties}}
The simulation results include the structural and electronic properties 
of clean SiC($0001$) and SiC($000 \overline{1}$) surfaces of all three polytypes. 
Following the results from the previous section in all cases the 12 DL slab models were
used to ensure good accuracy of the calculations. We also studied the dependence of the properties
on the different values of subsurface electric field, which was adjusted 
according to our findings described in Sec. \ref{Sec:SurfaceModels}.

\subsection{Structural properties}
\begin{figure}
\begin{center}
\includegraphics[width=0.7 \textwidth]{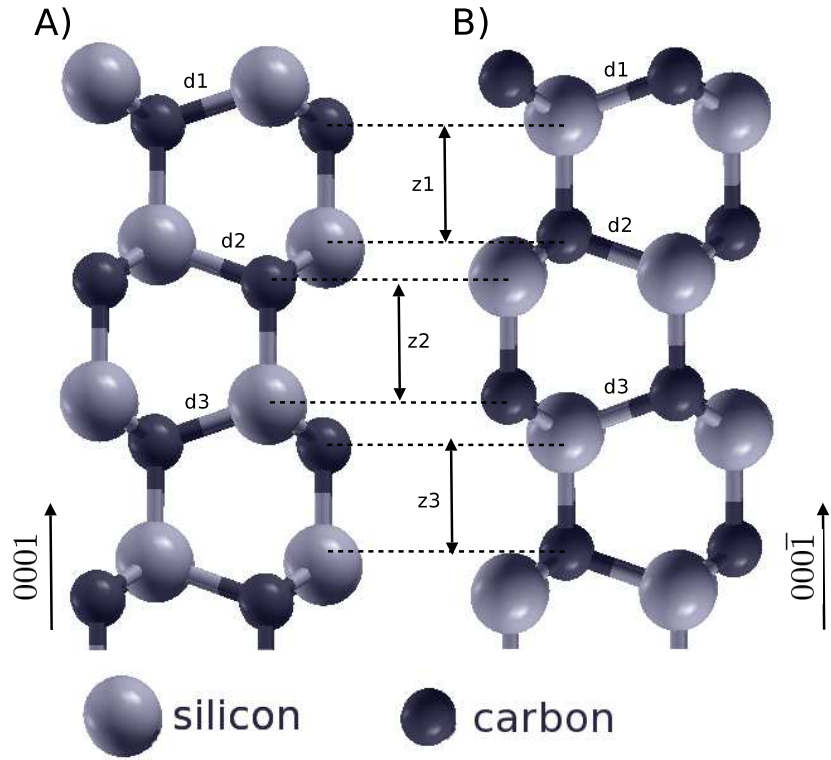}
\end{center}
\caption{\label{fig:relax} Side view of the relaxed 2H-SiC($0001$) and 2H-SiC($000\overline{1}$) surface. 
In this Figure Si atoms are represented by gray spheres and C atoms 
are by black spheres. Bond lengths related to relaxed atoms are showed in Table \ref{tab:table2} .}
\end{figure}

For bare SiC ($0001$) and SiC ($000 \overline{1}$) surfaces no surface reconstruction was obtained. 
The surface modification was limited to relaxation of several atomic layers (c.f. Fig. \ref{fig:relax}). 
The relaxation data is collected in Table \ref{tab:table2}.
\begin{table}
\caption{\label{tab:table2}
Relaxation of the upper part of 2H- , 4H- and 6H-SiC layers (in \AA)   }
\begin{center}
\begin{tabular}{c|ccc}
& \multicolumn{3}{c}{SiC($0001$) surface}  \\
\hline
 &  2H & 4H & 6H \\
\hline
$\Delta d1$ &   -0.0068 (-0.36 \% )& -0.0202 (-1.07 \%) & -0.0283 (-1.49 \%)         \\
$\Delta d2$ &   -0.0012 (-0.07 \% )& -0.0023 (-0.12 \%) & -0.0003 (-0.02 \%)     \\
$\Delta d3$ &   0.0002 (0.01 \% )& -0.0016 (-0.09 \%) & 0.0006 (0.03 \%)         \\
$\Delta z1$ &   0.0194 (1.02 \% )& 0.0374 (1.97 \%) & 0.0228 (1.20 \%)                \\
$\Delta z2$ &   -0.0006 (-0.03 \% )& 0.0087 (0.46 \%) & 0.0038 (0.20 \%)              \\
$\Delta z3$ &   -0.0005 (-0.03 \% )& -0.0016 (-0.08 \%) & 0.0045 (0.24 \%)       \\
\hline
\hline
&  \multicolumn{3}{c}{SiC($000 \overline{1}$) surface}  \\
\hline
 & 2H & 4H & 6H \\
\hline
$\Delta d1$ &  -0.0730 (-3.86 \% )& -0.0760 (-4.01 \%) & -0.0784 (-4.13 \%)        \\
$\Delta d2$ &  -0.0212 (-1.12 \% )& -0.0212 (-1.12 \%) & -0.0249 (-1.31 \%)         \\
$\Delta d3$ &  -0.0050 (-0.27 \% )& -0.0029 (-0.15 \%) & -0.0038 (-0.20 \%)     \\
$\Delta z1$ &  0.1027 ( 5.39 \% )& 0.1094 (5.75 \%) & 0.1108 (5.82 \%)              \\
$\Delta z2$ &  0.0233 ( 1.22 \% )& 0.0283 (1.49 \%) & 0.0194 (1.02 \%)               \\
$\Delta z3$ &  -0.0002 (-0.01 \% )& 0.0066 (0.35 \%) & 0.0050 (0.26 \%) 
\end{tabular}
\end{center}
\end{table}
From these data it follows that the significant relaxation is limited to the first two layers. 
The most significant changes are observed in the [$0001$] and [$000 \overline{1}$] direction.
No significant difference between the polytypes was observed. Also there is no dependence on 
the electric field which indicates weakly ionic character of the  polar SiC surfaces. 
The data is in good agreement with Ref. \cite{Sabish}. 

\subsection{Electronic properties}
\begin{figure*}
\includegraphics[width=\textwidth]{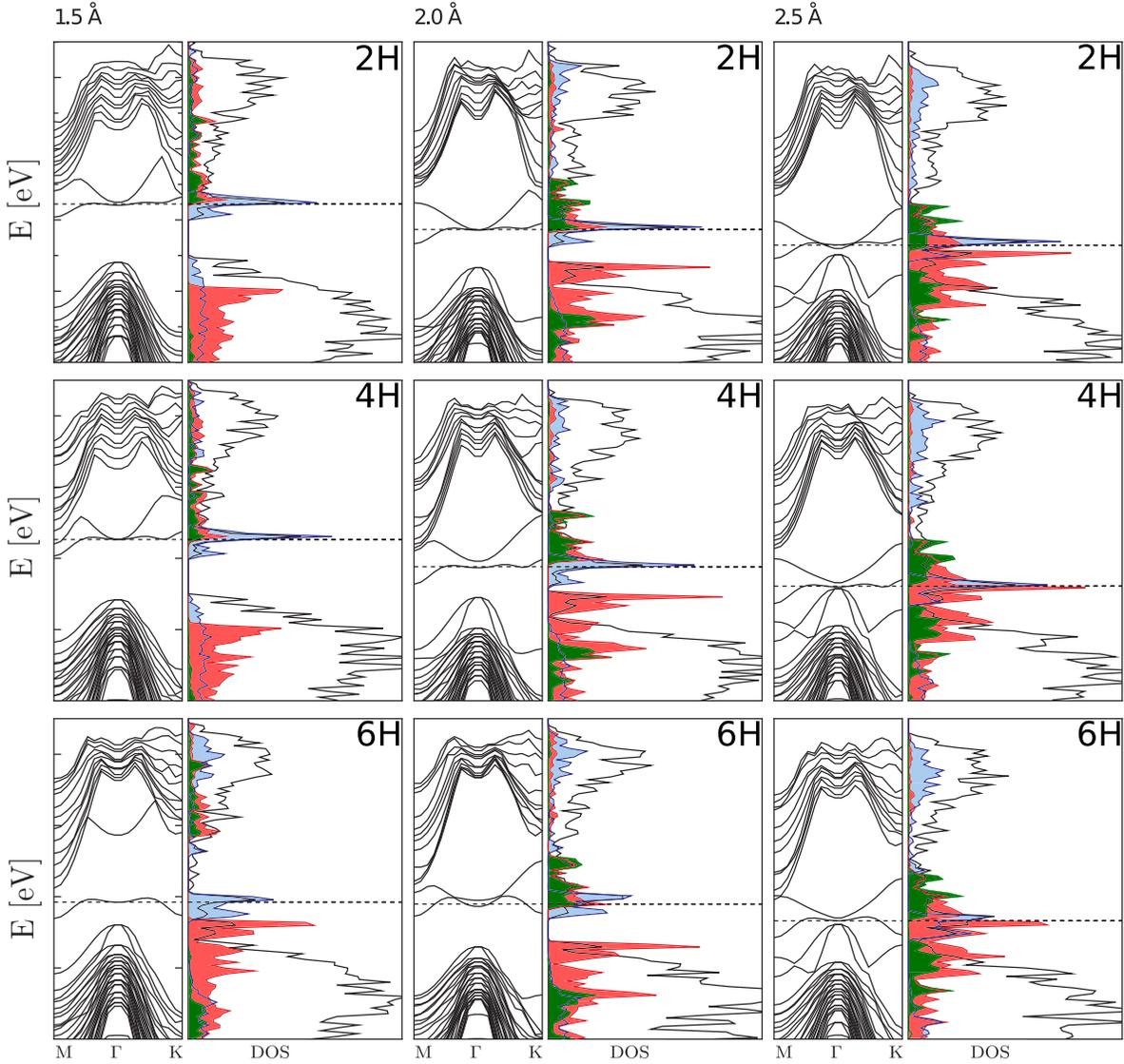}
\caption{\label{fig:banddoswithh} (Color online) (a) 2H-, (b) 4H- and (c) 6H-SiC($0001$) surface band diagram 
for the H-termination atoms located at the following distances for the bottom C atoms: 
1.5 \AA \  left, 2.0 \AA \  middle, and 2.5 \AA \  right. Density of states of H-termination (green), C-bottom (red)  
and Si-top (blue) atoms multiplied by four are presented in DOS plots. 
Fermi level is marked by dotted line. 
}
\end{figure*}

In contrast to structural properties, the electronic properties of bare SiC ($0001$) 
and SiC ($000 \overline{1}$) surfaces are severely affected by the subsurface electric field. 
This effect is shown in collection of diagrams plotted in Fig. \ref{fig:banddoswithh}. In Fig. \ref{fig:banddoswithh} the band diagram 
for the Si-surface of 2H polytype is presented.
These results indicate that the surface states associated with the Si atoms are moved down 
with respect to VBM. The two surface states are these associated with Si-top, C-bottom and 
hydrogen termination atoms. The band associated with the H atoms has considerable dispersion 
arising from the overlap with C dangling bonds. These states are essentially empty. 
The band related to Si-top atoms is almost dispersionless.  The energy of the states 
changes strongly with the electric field so that for these fields the states move across 
the entire bandgap. 
For 2H, 4H  two bands associated with bottom atoms (i.e. hydrogen and
carbon) and Si-top atoms are close to each other. 
Yet there are some differences between SiC polytypes. For 6H SiC($0001$) the states are initially
more separated  than for 2H, 4H. When the distance between hydrogen and carbon is increased the band 
related to the H- and C-bottom atoms moves to the Fermi level.
Fermi level moves in a valence band direction in all cases. 
Moreover, two additional bands associated with C-bottom atoms are shifted into valence band when 
the H-termination atoms are sufficiently close to the surface. 
With the diversion of hydrogen atoms case is more and more similar to the bare surface 
c.f. Fig. \ref{fig:banddos2h}.
In all these cases it is observed that the Fermi level is pinned to the surface 
states related to dangling Si bonds of the SiC(0001) surface.  

Field intensities have been shown in Tab. \ref{tab:table3}. For three polytypes (2H, 4H and 6H) 
electric field changes from positive to negative values c.f. Fig. \ref{fig:fieldwithh2h}.
However, for 6H-SiC (distance of 1.5 \AA) field is much smaller than for 2H- and 4H-SiC. 
%
\begin{table}
\caption{\label{tab:table3}
Field intensities [eV/\AA]  within 2H-, 4H- and 6H-SiC($0001$) slab deduced from the potential profiles.   }
\begin{center}
\begin{tabular}{cccc}
Distance [\AA] & 2H & 4H & 6H \\
\hline
1.5 &  0.054 &  0.049 &  0.035 \\
2.0 &  0.004 &  0.018 &  0.019 \\
2.5 & -0.020 & -0.026 & -0.030
\end{tabular}
\end{center}
\end{table}
\begin{figure*}[ht]
\includegraphics[width=\textwidth]{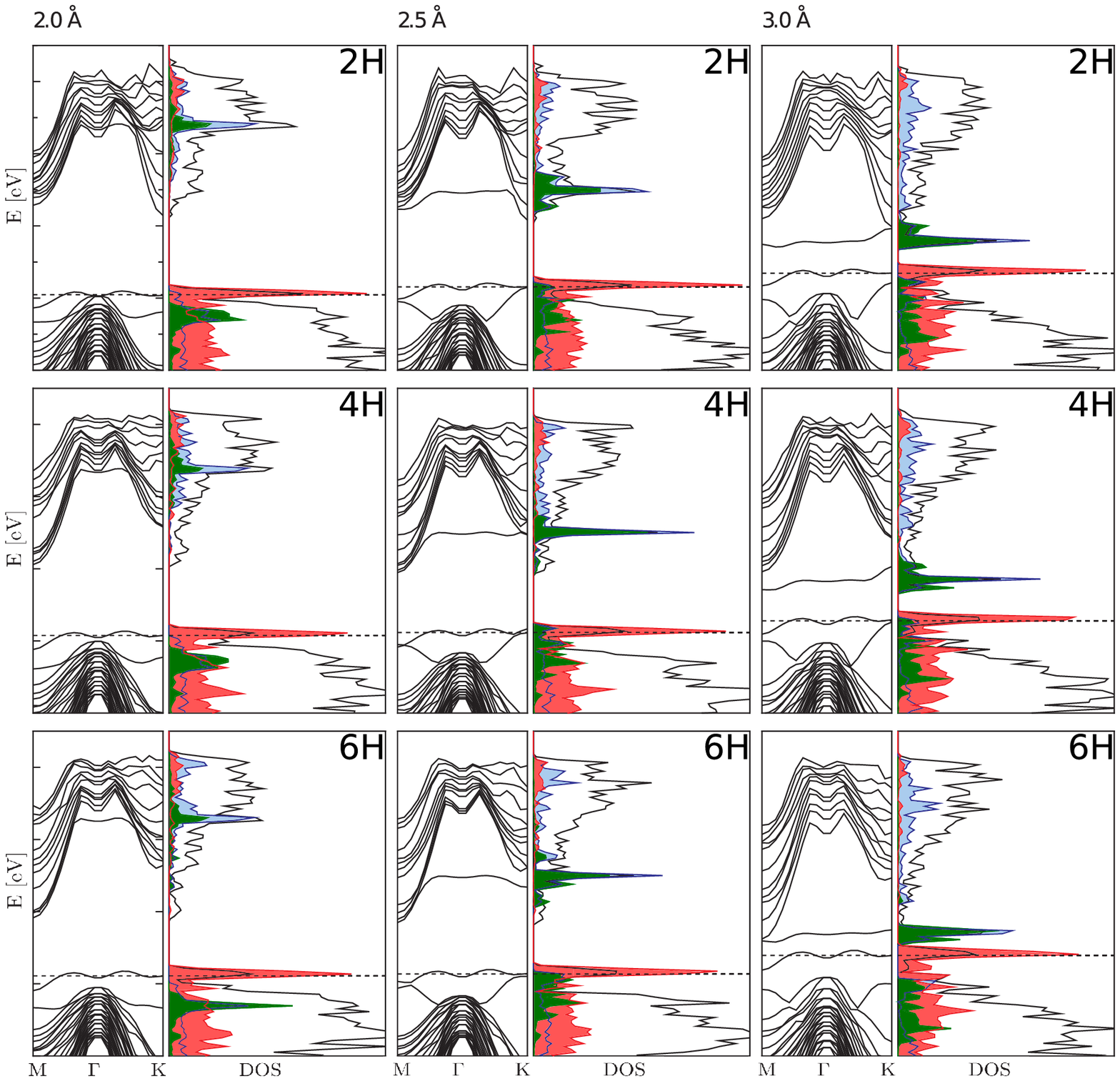}
\caption{\label{fig:banddoswithh2} (Color online) (a) 2H-, (b) 4H- and (c) 6H-SiC($000 \overline{1}$) surface band diagram
for the H-termination atoms located at the following distances for the bottom Si atoms:
2.0 \AA \  left, 2.5 \AA \  middle, and 3.0 \AA \  right. Density of states of H-termination (green), Si-bottom (blue)
and C-top (red) atoms multiplied by four are presented in DOS plots.
Fermi level is marked by dotted line.
}
\end{figure*}

Similar analysis was made for C-terminated SiC($000 \overline{1}$). Band structures and DOS are presented in Fig. \ref{fig:banddoswithh2}.
As in the previous case overlap of H-terminated and Si-bottom wave function strongly affects electronic structure. 
Unlike SiC($0001$) surface electric field changes from negative to positive values.  
Band associated with C-top atom states initially are located near the VBM, but under influence of electric field 
are moved across the entire band gap. It is dispersionless.  
On the other hand, band associated with bottom atoms states (H- and Si-bottom) resemble the shape of the valence bands. 
However, when the hydrogen atoms are moved, it bends to the Fermi level. 
Additionally, there is one more band (also associated with bottom atoms states) shifted outside of the conduction band,
i.e., with a change of electric field within the slab it is moved to the Fermi level.
Moreover, dispersion of this band disappears with increasing distance of H atoms.
Fermi level is pinned to the termination atom states.

The surface states are divided into Shockley \cite{Shockley} and Tamm \cite{Tamm} states. It is often claimed that there is no real physical 
distinction between these two types. In fact there is deep difference between them. Shockley surface states are 
those associated with band states which are localized at the surface by the potential well created by band bending (see Fig. \ref{fig:acpdon}). 
Thus they disappear when there is no potential well. They are localized only in the direction perpendicular to the surface. 
Tamm states of surface atoms arise from dangling bonds at the surface. They are fully localized and they 
do not play role in the electrical conductivity. 
In contrast to Shockley states, Tamm states do not disappear with the potential well. 
This difference provides clear procedure for determination of the
type of surface states.
Generally, it is difficult to precisely indicate which states are Shockley and Tamm states. 
There are several works in recent literature describing way in which this states can be 
distinguished \cite{Zak}.
Zak showed graphical description of two kinds of crystal terminations. For different termination types  we expect
Shockley or Tamm states. Following this classification surface states induced by surface perturbation 
should be called Tamm states. 
In Fig. \ref{fig:field2h} we can observe such perturbation at the edge of 
each bare surface, i.e., shallower wells at the ends of the slabs. Because of this perturbation 
Tamm surface states can be induced. 

Moreover, it was found that Tamm and Shockley surface states can coexist in semiconducting superlattice \cite{Klos}. 
To denote type of states band structure and DOS have been analyzed 
for different methods of termination. The states which disappear in the reversed electric field are 
Shockley states. The states which move across the entire band gap preserving their localization 
at the particular surface are Tamm states. 

Following changes in the band structure and DOS and comparing results in SiC($0001$) and SiC($000 \overline{1})$ side
it is possible to identify surface states associated with Si and C edge atoms.
This method can be used as a criterion for identification of Tamm and Shockley surface states.  
 
Moreover, changing position of H-termination atoms we can move Fermi level 
which means that we can control doping type of SiC surface. 
In Fig. \ref{fig:banddoswithh} and  Fig. \ref{fig:banddoswithh2} doping type is changed from n-type and p-type 
respectively to almost undoped SiC surface.

%
\begin{table}
\caption{\label{tab:table4}
Field intensities [eV/\AA]  within 2H-, 4H- and 6H-SiC($000 \overline{1}$) slab deduced from the potential profiles.   }
\begin{center}
\begin{tabular}{cccc}
Distance [\AA] & 2H & 4H & 6H \\
\hline
2.0 & -0.026 & -0.024 & -0.031 \\
2.5 &  0.006 &  0.010 &  -0.014 \\
3.0 &  0.047 &  0.033 &  0.066 
\end{tabular}
\end{center}
\end{table}
\section{Summary\label{Sec:Summary}}
Extensive density functional study of bare and H-terminated SiC surfaces 
was presented. Different SiC($0001$) and SiC($000 \overline{1}$) polytypes: 
2H, 4H, 6H were considered and compared. It was determined that twelve Si-C layers
is a sufficient number for high precision calculations.    
Structural optimization was performed for 2H- , 4H- and 6H- 
SiC($0001$) as well as SiC($000 \overline{1}$) surface and no significant  
differences between them were found. 
It was shown that for all cases (H-terminated 2H-, 4H-, 6H-SiC) electric field within 
the slab can be adjusted. Field intensity dependence on the H positions 
for SiC($0001$) and SiC($000 \overline{1}$) have 
been presented (Table \ref{tab:table3} and \ref{tab:table4}).
Figures \ref{fig:banddoswithh} and \ref{fig:banddoswithh2}
prove that Fermi level is pinned to the surface states.
The energy of surface related states of both Si- and C-terminated 
surfaces are moved by the subsurface field with respect to VBM. 
In case of Si-terminated surface the energy is shifted by about 
2 eV while in the C-terminated surface, this shift amounts to about 1 eV. 
This effect can be used as a simulation tool, e.g., for 
the identification and subsequent analysis of surface states as well as 
changing n-type and p-type doping level.

\section*{Acknowledgements}
The calculations reported in this paper were performed using computing 
facilities of the Interdisciplinary Center for Mathematical and Computational Modeling (ICM) of the University of Warsaw. 
The research was supported by the European Union within European 
Regional Development Fund, through grant Innovative Economy 
(POIG.01.01.02-00-008/08).


\section*{References}
\begin{thebibliography}{36}
\bibitem{Kackell} F.K{\"a}ckell, J. Furthm{\"u}ller and F.Bechstedt, {\it Phys. \ Rev. B 60, 13261 (1997)}
\bibitem{Masri} P.Masri, {\it Surf. \ Sci. \ Rep. 48,1 (2002)}
\bibitem{Bechstedt} F. Bechstedt and F. K{\"a}ckell {\it Phys. \ Rev. \ Lett. 75, 2180 (1995)}
\bibitem{ru} {\it see e.g. http://www.ioffe.ru/SVA/NSM/Semicond/SiC/}
\bibitem{Tang}L. Tang, M. Mehregany and L. Matus, {\it Appl. \ Phys.\ Lett. 60, 2992 (1992)}
\bibitem{Goldberg} Y. Goldberg, M. Levinshtein and S. Rumyantsev {\it in Properties of Advanced Semiconductor Materials GaN AlN SiC BN SiC SiGe . Eds. Levinshtein M.E. Rumyantsev S.L. Shur M.S. John Wiley Sons, Inc., New York, pp. 93-148 (2001)}
\bibitem{Capano} M. Capano and R. Trew {\it Mater. Res. Soc. Bull. 22, 19 (1997)}
\bibitem{Tairov1} Y.~M. Tairov. and V. Tsvekov, {\it J. Cryst. Growth 43, 209 (1978)}
\bibitem{Tairov2} Y.~M. Tairov and V. Tsvekov, {\it J. Cryst. Growth 52, 146 (1981)}
\bibitem{Verma} A.~R. Verma {\it Crystal Growth and Dislocations, Academic Press, New York Butterworths, London (1953)}
\bibitem{Sabish} M. Sabisch, P. Kr{\"u}ger and J.Pollmann, {\it Phys. Rev. B 55, 10561 (1997)}
\bibitem{Catellani} A. Catellani and G. Galli {\it Prog. Surf. Sci. 69, 101(2002)}
\bibitem{Starke} U. Starke, J. Schardt, J. Bernhardt, M. Franke, K. Reuter, H. Wendler, K. Heinz, J. Furthmuller, P. K{\"a}ckell and F. Bechstedt, {\it Phys. Rev. Lett. 80, 758 (1998)}
\bibitem{Kulakov} M.~A. Kulakov, G. Henn and B. Bullemer {\it Surf. Sci. 49, 346 (1996)}
\bibitem{Li} Y.~L.~L. Xe and X. Wang, {\it Surf. Sci. 600, 298 (1996)}
\bibitem{Hass} J. Hass, W.~A. de~Heer and E. Conrad,{\it J. Phys.: Condens. Matter 20, 323202 (2008)}
\bibitem{Righi}M. Righi, C. Pignedoli, G. Borghi, R.~D. Felice, C. Bertoni and A. Catellani, {\it Phys. Rev. B 66, 45320 (2002)}
\bibitem{Kresse1993} G.Kresse and J.Hafner, {\it Phys. Rev. B 47,558 (1993)}
\bibitem{Kresse1996}G. Kresse and J. Furthm{\"u}ller,{\it Comput. Mat. Sci.6,15 (1996)}
\bibitem{Kresse1996a}G. Kresse and J. Furthm{\"u}ller, {\it Phys. Rev. B 54, 11169 (1996)}
\bibitem{Blochl} P.~E. Bl{\"o}chl, {\it Phys. Rev. B 50, 1793 (1994)}
\bibitem{Kresse1999} G. Kresse and D. Joubert,{\it Phys. Rev. B 59,1758 (1999)}
\bibitem{Perdew1996} J.Perdew, K.Burke and M.Ernzerhof, {\it Phys. Rev. Lett. 77, 3865 (1996)}
\bibitem{Monkhorst} H.~J. Monkhorst and J.~D. Pack, {\it Phys. Rev. B 13, 5188 (1976)}
\bibitem{Schulz} H.Schulz and K.Thiemann {\it Solid State Communications 32, 9 (1979)}
\bibitem{Emtsev} K.~V. Emtsev, T. Seyller, L. Ley, L. Broekman, A. Tadich, J.~D. Riley, R.~G.~C. Leckey, and M. Preuss, {\it Phys. Rev. B 73, 075412 (2006)}
\bibitem{Neugebauer} J. Neugebauer and M. Scheffler, {\it Phys. Rev. B 46,16067 (1992)}
\bibitem{Krukowski1}S. Krukowski,P. Kempisty and P. Str\c{a}k, {\it J. Appl. Phys. 105, 113701 (2009)}
\bibitem{Krukowski2} S. Krukowski, P. Kempisty, P. Str\c{a}k and K. Sakowski, {\it J. Appl. Phys. 106, 054901 (2009)}
\bibitem{Meyer} B. Meyer and D. Marx, {\it Phys. Rev. B 67, 35403 (2003)}
\bibitem{Tasker} P.~W.Tasker, {\it J. Phys. C 12, 4977 (1979)}
\bibitem{Noguera}C. Noguera, {\it J. Phys.: Condens. Matter 12, R367 (2007)}
\bibitem{Shockley} W. Shockley, {\it Phys. Rev. 56 ,317 (1939)}
\bibitem{Tamm} I. Tamm, {\it Phys. Z. Soviet Union 1, 733 (1932)}
\bibitem{Zak} J. Zak, {\it Phys. Rev. B 32, 2218 (1985)}
\bibitem{Klos} J. K\l{}os and H. Puszkarski, {\it Phys. Rev. B 68, 045316 (2003)}
\end{thebibliography}
\end{document}